\documentstyle[12pt]{article}

\title{The Strong Perturbation Approach for the Dirac Equation in a Gauge
Field}

\author{Marco Frasca \\
Via E. Gattamelata,3 00176 Roma (Italia) \\
e-mail: mc3747@mclink.it
}

\date{17 June 1994}

\maketitle

\begin{abstract}

We discuss a possible approach to the problem of a gauge theory with a
strong coupling constant. It is seen that, instead of plane waves, we
have to consider the adiabatic eigenstates of the perturbation in order
to get a meaningful perturbation approach.

\end{abstract}

\newpage

\begin{document}

We consider the following model of a fermion in a gauge field[1]
\begin{equation}
    i\frac{\partial\psi}{\partial t} =          \label{eq:dirac}
    -i\mbox{\boldmath$\alpha$}\cdot\nabla\psi + \beta m\psi
    -g\mbox{\boldmath$\alpha$}\cdot{\bf A}_i\lambda_i\psi
    + g A^0_i\lambda_i\psi
\end{equation}
having set $\hbar=c=1$ and $\mbox{\boldmath$\alpha$}$ and $\beta$ the
Dirac
matrices. ${\bf A}_i$ and $A^0_i$ are the gauge fields with generators
$\lambda_i$. The coupling constant is $g \gg 1$. In the following we put
\begin{equation}
    H_0 = -i\mbox{\boldmath$\alpha$}\cdot\nabla + \beta m
\end{equation}
the unperturbed hamiltonian, and
\begin{equation}
    V = -\mbox{\boldmath$\alpha$}\cdot{\bf A}_i\lambda_i
    + A^0_i\lambda_i
\end{equation}
the perturbation.

To approach this problem we consider a path similar to the one followend
in
Ref.[2]. Then we consider the equation
\begin{equation}
    i\frac{\partial U}{\partial t} = g V U  \label{eq:eqU}
\end{equation}
being $U$ a time evolution operator so that we can put
\begin{equation}
    \psi = U \tilde{\psi}
\end{equation}
with $\tilde{\psi}$ satisfying the equation
\begin{equation}
    i\frac{\partial\tilde{\psi}}{\partial t} = U^{-1} H_0 U \tilde{\psi}
\end{equation}
from which we derive the following perturbation series
\begin{equation}
    \tilde{\psi}(t) = T\exp\left[
                             -i\int^t_{t_0} dt' U^{-1}(t') H_0 U(t')
                           \right]\psi(0)
\end{equation}
remembering that $\tilde{\psi}(0) = \psi(0)$. So, it would be interesting
to
be able to get a solution for eq.(\ref{eq:eqU}) in the limit of $g \gg 1$.

This problem can be solved if we introduce the eigenstates of the
perturbation
$V(t)$, indicated by $|n; t>$ that satisfy the eigenvalue problem
\begin{equation}
    V(t) |n; t> = \epsilon_n(t) |n; t>.
\end{equation}
Next, we consider the start of the perturbation series for $\psi$ as given
by
$\psi^{(0)}(t) = U(t)\psi(0)$, so what is the form of the latter in the
limit
$g \gg 1$? We put
\begin{equation}
    \psi^{(0)}(t) = \sum_n a_n(t) e^{i\gamma_n(t)}
                                  e^{-i g \int^t_{t_0} dt' \epsilon_n(t')}
                                  |n; t>
\end{equation}
with $a_n(t)$ given by the exact equations
\begin{equation}
    i\dot{a}_m(t) = -\sum_{n \neq m} a_n(t) \label{eq:an}
                          e^{i[\gamma_n(t)-\gamma_m(t)]}
                          e^{-i g \int^t_{t_0} dt' \epsilon_n(t')
                          -\epsilon_m(t')}
                          <m; t|i\frac{\partial}{\partial t}|n; t>
\end{equation}
with
\begin{equation}
    \dot{\gamma}_m = <m; t|i\frac{\partial}{\partial t}|m; t>
\end{equation}
the geometrical phase. When we write eq.(\ref{eq:an}) in integral form, it
is
quite easy to realize that, in the limit of $g \gg 1$, for a well-known
theorem
of asymptotics[3], we have
\begin{equation}
    a_n(t) = a_n(t_0) + O(\frac{1}{\sqrt{g}})
\end{equation}
for crossing of the eigenvalues at some t or best, by integrating by
parts,
\begin{equation}
    a_n(t) = a_n(t_0) + O(\frac{1}{g})
\end{equation}
that are the results we were looking for.

In conclusion we have that the problem of a fermion in a gauge theory with
a strong coupling constant can be approached using both the result of
Ref.[2]
and the long time known results of the adiabatic theory [4].

REFERENCES

[1] M.Kaku, {\sl Quantum Field Theory}, (Oxford University Press, New
York,
1993)

[2] M.Frasca, Phys. Rev. A {\bf 45}, 43 (1992)

[3] R.B.Dingle, {\sl Asymptotic Expansions: Their Derivation and
Interpretation}, (Academic Press, London, 1973)

[4] M.Born, Z. Physik {\bf 40}, 167 (1926)

\end{document}